\documentclass[12pt]{article}
\pdfoutput=1
\input epsf.sty
\usepackage{hyperref}
\hypersetup {
	colorlinks=true,
	linkcolor=blue,
	linktocpage=true,
	citecolor=blue,
	urlcolor=blue
}
\usepackage{geometry}
\usepackage[utf8]{inputenc}

\usepackage{amsmath}
\usepackage{amsfonts}
\usepackage{amssymb}
\usepackage{cite}
\usepackage{graphicx}

\newcommand{\mIm}{\text{Im}\,}
\newcommand{\mRe}{\text{Re}\,}

\numberwithin{equation}{section}

\newcommand{\be}{\begin{equation}}
\newcommand{\ee}{\end{equation}}
\newcommand{\ben}{\begin{eqnarray}\displaystyle}
\newcommand{\een}{\end{eqnarray}}

\def\sqr#1#2{{\vcenter{\vbox{\hrule height.#2pt
         \hbox{\vrule width.#2pt height#1pt \kern#1pt
            \vrule width.#2pt}
         \hrule height.#2pt}}}}

\begin{document}

\begin{center}
{
\Large{\bf A STRINGY GLIMPSE INTO THE \\\vspace{5mm} BLACK HOLE HORIZON  }}

\vspace{10mm}


\textit{ Nissan Itzhaki and Lior Liram}
\break

Physics Department, Tel-Aviv University, \\
Ramat-Aviv, 69978, Israel \\


\end{center}

\vspace{10mm}

\begin{abstract}

We elaborate on the recent claim  \cite{Ben-Israel:2017zyi} that non-perturbative effects in $\alpha'$, which are at the core of the FZZ duality, render the region just behind the horizon of the  $SL(2,\mathbb{R})_k/U(1)$ black hole singular already at the  classical level ($g_s=0$). We argue that the 2D {\em classical}  $SL(2,\mathbb{R})_k/U(1)$ black hole could shed some light on  {\em quantum} black holes in higher dimensions including   large black holes in $AdS_5\times S^5$.

\end{abstract}

\baselineskip=18pt

\newpage

\section{Introduction}

In General Relativity the Black Hole (BH) horizon is a smooth region that  an infalling observer can safely cross. Quantum mechanically, however, the situation is potentially more complex as there are several arguments that support the existence of a non-trivial structure at the horizon of quantum black holes \cite{Itzhaki:1996jt, Braunstein:2009my,Mathur:2009hf,Almheiri:2012rt,Marolf:2013dba,Polchinski:2015cea}

None of the papers mentioned above explain the origin of the structure at the quantum BH horizon. Instead they argue, in different ways, that if the information is emitted with the radiation then the horizon cannot be smooth.  If indeed this is the case and if, as suggested by the AdS/CFT correspondence, the radiation contains the information then the challenge is to understand the nature and origin of the structure at the horizon. This  seems to be an extremely difficult challenge that appears to involve understanding quantum gravity at the full non-perturbative level.

Recently \cite{Ben-Israel:2017zyi}, a surprising development took place that could potentially help in this regard. Taking advantage of the fact that the $SL(2,\mathbb{R})_k/U(1)$ BH model is  solvable on the sphere ($g_s=0$) and that, in particular, the reflection coefficient is known exactly \cite{Teschner:1999ug}, it was argued  that a non-trivial structure just behind the horizon of the $SL(2,\mathbb{R})_k/U(1)$
BH appears already at the {\em classical} level ($g_s=0$).

To a large extent, the claim made in \cite{Ben-Israel:2017zyi} is the Lorentzian analog of \cite{Giveon:2015cma} where the Euclidean version of the $SL(2,\mathbb{R})_k/U(1)$ BH was studied. It was argued that the FZZ duality \cite{fzz,Kazakov:2000pm} implies that due to classical, non-perturbative $\alpha'$  effects, high energy modes see a different geometry than low energy ones do. Compared with the classical (SUGRA) background, the Euclidean horizon (i.e. the tip of the cigar) seems to be modified the most. This was further studied in \cite{Ben-Israel:2015mda} where it was shown that similar conclusions regarding the tip of the cigar could be made by directly analyzing the reflection coefficient, without employing the FZZ duality. This serves as a motivation to study the Lorentzian BH for which, at present, no known analog of the FZZ duality exists.

The fact that the {\it classical} $SL(2,\mathbb{R})_k/U(1)$ BH seems to have structure at the horizon, suggests that this model captures some aspects of horizons of higher dimensional {\it quantum} BHs.
The  goal of the present note is  to further explore this exciting possibility and to elaborate on the results of \cite{Ben-Israel:2017zyi}. In particular, we take advantage of the, somewhat hidden,  symmetries associated with  scattering in the BH background, that were ignored in \cite{Ben-Israel:2017zyi}, to shed light on  some issues associated with the results of \cite{Ben-Israel:2017zyi}.  We also clarify the sense in which  the 2D classical $SL(2,\mathbb{R})_k/U(1)$ BH is related to some quantum BHs in higher dimensions that include large BHs in $AdS_5\times S^5$.

The paper is organized as follows. In the next section we point out  a relationship between the BH singularity and  the reflection coefficient of a wave that scatters on the BH at high energies. We also illustrate this relationship for
the Schwarzschild BH. In section 3 we consider this relationship in the case of the $SL(2,\mathbb{R})_k/U(1)$ BH at the SUGRA level.  In this case, both the background and the reflection coefficient are known exactly.
We show that they fit neatly with the relationship of section 2.  In section 4 we study the $SL(2,\mathbb{R})_k/U(1)$ BH including perturbative corrections in $\alpha'$. Here too, both the background and the reflection coefficient are known exactly and we show that they agree with the relationship of section 2. In section 5 we consider the exact $SL(2,\mathbb{R})_k/U(1)$ BH on the sphere ($g_s=0$), including non-perturbative corrections in $\alpha'$. The reflection coefficient is known exactly, but the background is not. We use the results of section 2 to learn about the structure of the singularity in this case. We  find that the non-perturbative corrections in $\alpha'$ push the singularity all the way to the horizon. Section 6 is devoted to discussion.

\section{The reflection coefficient and the BH singularity}

In this section we point out a relation between the BH singularity and the reflection coefficient at high energies.

When probing a target with a wave it is standard that the S-matrix elements at high energies are most sensitive to the singular features of the potential associated with the target. There are (at least) two reasons {\em not} to expect this to hold for BHs:
\begin{enumerate}
\item As we increase the energy most of the wave gets absorbed by the BH and there appears to be little information outside the BH that can be used to probe the BH singularity.
\item The BH singularity is surrounded by the horizon. Causality implies that a wave that probes the singularity cannot escape back to infinity and, in particular, it cannot contribute to the reflection coefficient.
\end{enumerate}

Point (1) above makes it hard to imagine that there is a relation between the BH singularity and the reflection coefficient at high energies and it seems that point (2) makes it impossible. Nevertheless, we shall see that such a relationship does exist due to a (hidden) symmetry in the scattering problem associated with the BH.

When considering the problem of  scattering  a wave off a BH it is useful to use the tortoise coordinate $x$, along with the usual Schwarzschild time $t$. They are related to Kruskal coordinates in the following way
\begin{equation}
t =\frac{\beta}{4\pi}\log(U/V), ~~~~~
x = \frac{\beta}{4\pi}\log(U V) \label{x}.
\end{equation}

The tortoise coordinate covers the BH exterior such that in the asymptotic region, $x=\infty$, and at the horizon, $x=-\infty$.
What is particularly nice about  the tortoise coordinates is that the wave equation takes the form of a Schr\"{o}dinger equation:
\begin{equation}\label{schrodingerEquation}
\left(-\frac{d^2}{d x^2}  + V(x) -\omega^2 \right) \psi(x) = 0,
\end{equation}
where $\omega$ is the energy (conjugate to $t$) of the mode.
Therefore, the relation between $V(x)$ and the reflection coefficient, $R(\omega)$, is the usual one known from quantum mechanics. Consequently, we use the Born approximation to compute $R(\omega)$ at high energies, as described below. 

\begin{figure}
\centerline{\includegraphics[width=0.7\textwidth]{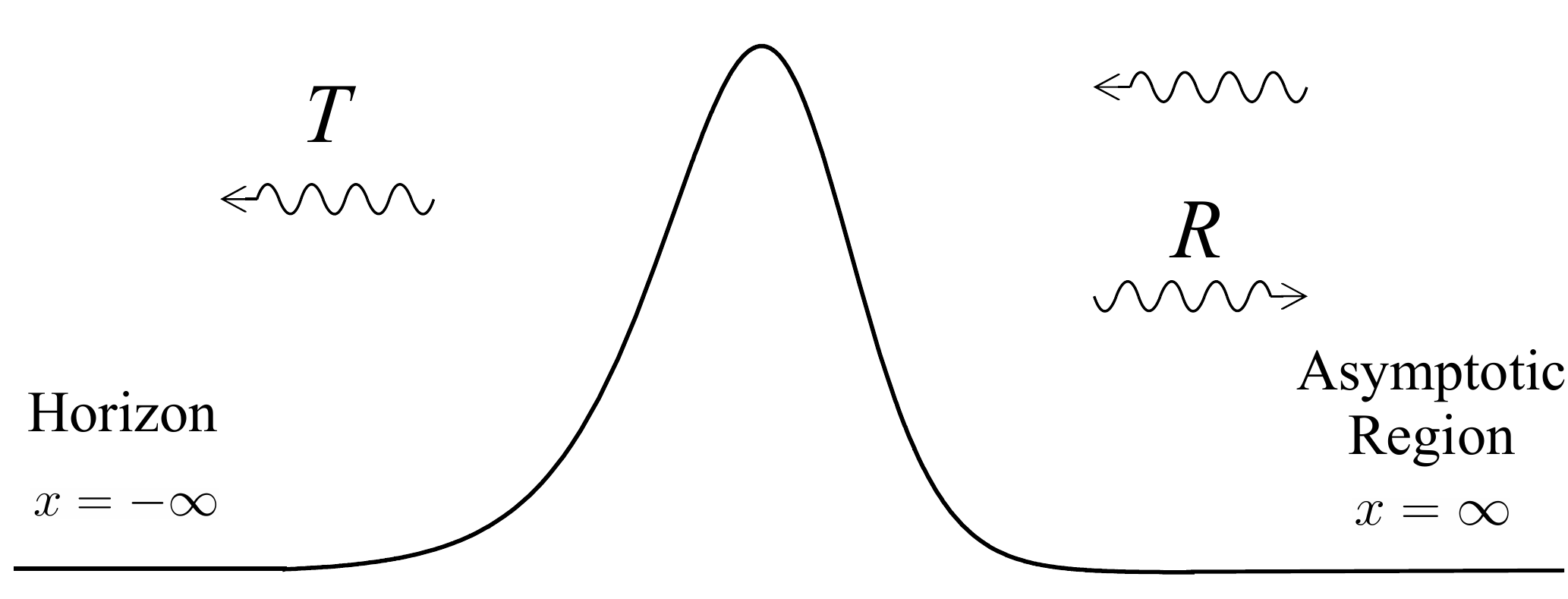}}
\caption{BH Scattering in tortoise coordinates. A wave arriving from the asymptotic region ($x=\infty$) is scattered off the potential. At high energies, most of the wave reaches the horizon ($x=-\infty$) and absorbed by the BH, while a small fraction ($R \ll 1$) is reflected back.}
\label{fig:scatter}
\end{figure}

We consider a process where an incoming wave arrives from $x=\infty$. At high energies most of it gets across the potential,  reaches $x=-\infty$ and gets absorbed by the BH. Some of it is reflected back to $x=\infty$ (see figure \ref{fig:scatter}). This emphasizes the claim that the reflection coefficient cannot be sensitive to the BH singularity as the whole process happens outside the BH.

\begin{figure}
\centerline{\includegraphics[width=0.75\textwidth]{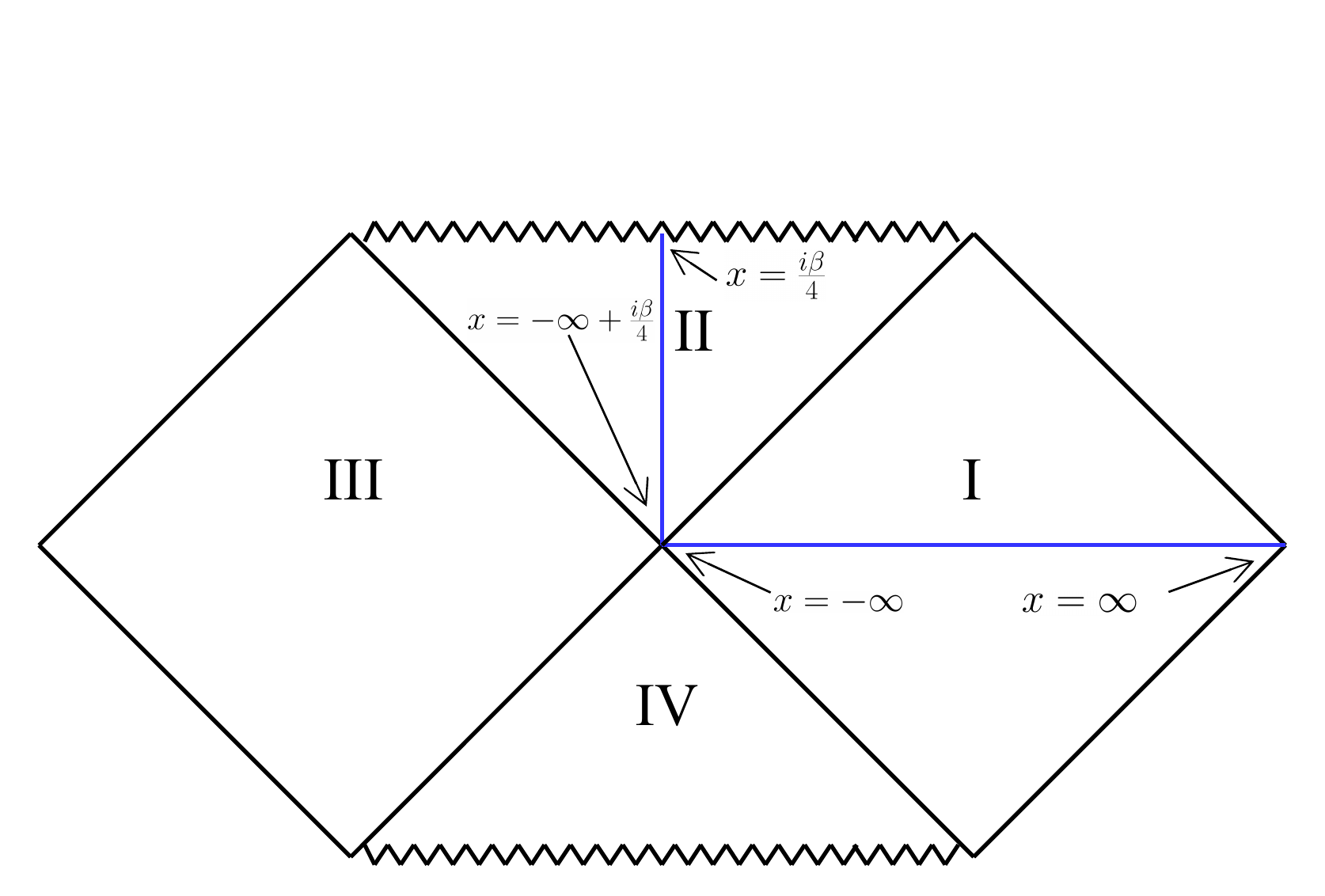}}
\caption{Penrose diagram of a Schwarzschild BH. To reach region II, we shift the tortoise coordinates: $x \to x + i\beta /4,~ t \to t + i\beta /4$.
The blue lines are the $t=0$ slices in regions I \& II. The horizon is mapped to $x=-\infty$ in both regions. }
\label{fig:Penrose}
\end{figure}

There is a symmetry in this scattering process that turns out to be quite powerful.  Eq. (\ref{x}) imply that points related by
\be
t\to t+\frac{i \beta}{2}(n+m),~~~x\to x+\frac{i \beta}{2}(n-m),~~~\mbox{where}~~~~n,m\in Z
\ee
should be identified. If we divide the maximally extended manifold into regions in the usual way (see figure \ref{fig:Penrose}), then we see that keeping $t$ intact while taking $x\to x \pm i \beta$ keeps us in the same region. Therefore, the potential $V(x)$ is periodic in imaginary $x$,
\be
V(x\pm i \beta)=V(x).
\ee
Keeping $t$ intact while taking $x\to x \pm i \beta/2$ takes a point in region I to its mirror in region III. The potential in region III is identical to the potential in region I, hence the periodicity is in fact
\be\label{shift}
V(x\pm i \beta/2)=V(x).
\ee
Taking $x\to x \pm i \beta/4$, while $t \to t \pm i \beta/4$ takes a point in region I to a point in region II (and IV). This fact and the periodicity of the potential will play a key role in what follows.

Going back to the scattering process, at high energies (compared to $1/\beta$, that is the scale in the problem) we can use the Born approximation in 1D to calculate
the reflection coefficient,
\begin{equation}\label{BornApproximation}
R(p)  = \frac{1}{2 i p} \int_{-\infty}^{\infty} dx \, V(x) e^{-2 i p x},
\end{equation}
with $p$ the momentum which we take to be equal to the energy $\omega$ (i.e. we consider massless fields).

\begin{figure}
\centerline{\includegraphics[width=0.75\textwidth]{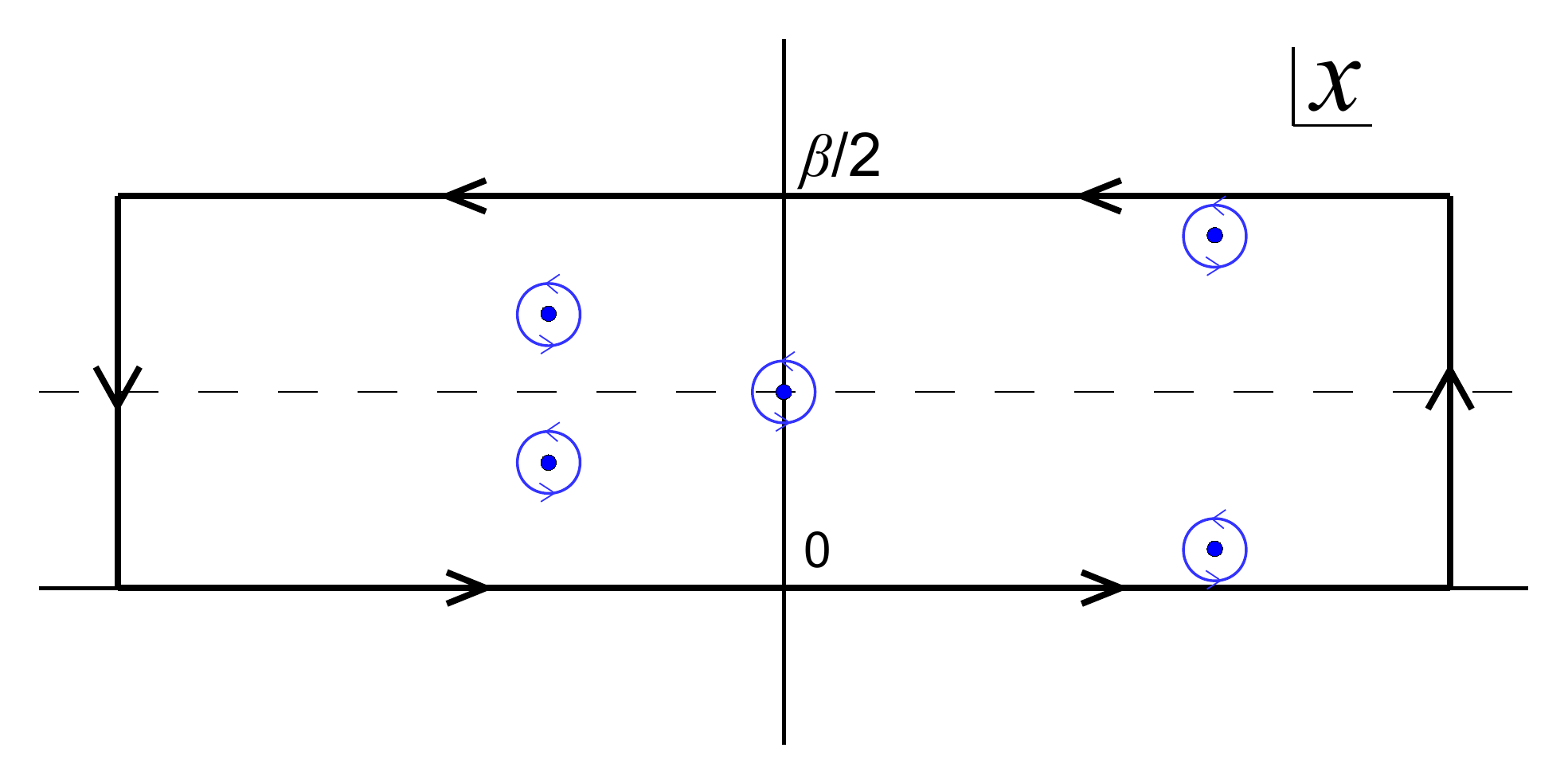}}
\caption{The contour used to calculate the reflection coefficient in the Born approximation. It goes counter clockwise around the strip $|\mRe{x}| \leq \infty,~ 0\leq \mIm{x} \leq \beta / 2$. Blue dots represent possible locations of poles of $V(x)$. The large contour (black) is equivalent to the sum of the small contours (blue) encircling the poles.}
\label{fig:baseContour}
\end{figure}

Instead of performing the integration from $x=-\infty$ to $x=\infty$, let us consider the integral in the complex $x$-plane
$$\int_{C} dx V(x)e^{-2 i p x}$$ along a contour, $C$, that we now specify. Let us start with the simplest case in which $V(x)$ has no branch points. This case  is relevant for the setup considered in next section. In this case we take the contour $C$ to be as plotted in figure \ref{fig:baseContour}. The contribution to the integral of the two vertical lines vanish since the potential vanishes at $x=\pm \infty.$ Because of the symmetry (\ref{shift}) the contribution of the two horizontal lines is identical up to a factor of $e^{\beta p}$. With the help of the residue theorem we get,
\be\label{ju}
R(p)(1-e^{\beta p})=\frac{\pi}{p} \sum \mbox{Res} \left(V(x) e^{-2ipx}\right),
\ee
where the sum runs over the poles of $V(x)$ within the rectangle.

Since the poles in $V(x)$ are related to   the BH singularity,  (\ref{ju}) relates the reflection coefficient at high energies with the BH singularity.
Roughly speaking, what happened is that the symmetry (\ref{shift}) allowed us to surround the singularity (in the complex plane) and by doing so  evading the arguments from the beginning of the section that explain why there can be no relation between the singularity and the reflection coefficient.

We are interested in the high energy limit. In that limit we can ignore the factor of `1' on the left hand side of (\ref{ju}) and the main contribution to the right hand side comes from the poles closest to the line $\mIm(x)= \beta /2$. We end up with
\be\label{uj}
R(p)=  \frac{\pi }{p}e^{-\beta p} \sum_{\substack{\text{\fontsize{10pt}{10pt}\selectfont closest}
 \\ \text{\fontsize{10pt}{10pt}\selectfont poles}}} \mbox{Res} \left(V(x) e^{-2ipx}\right),
\ee
at the UV.

\begin{figure}
\centerline{\includegraphics[width=0.75\textwidth]{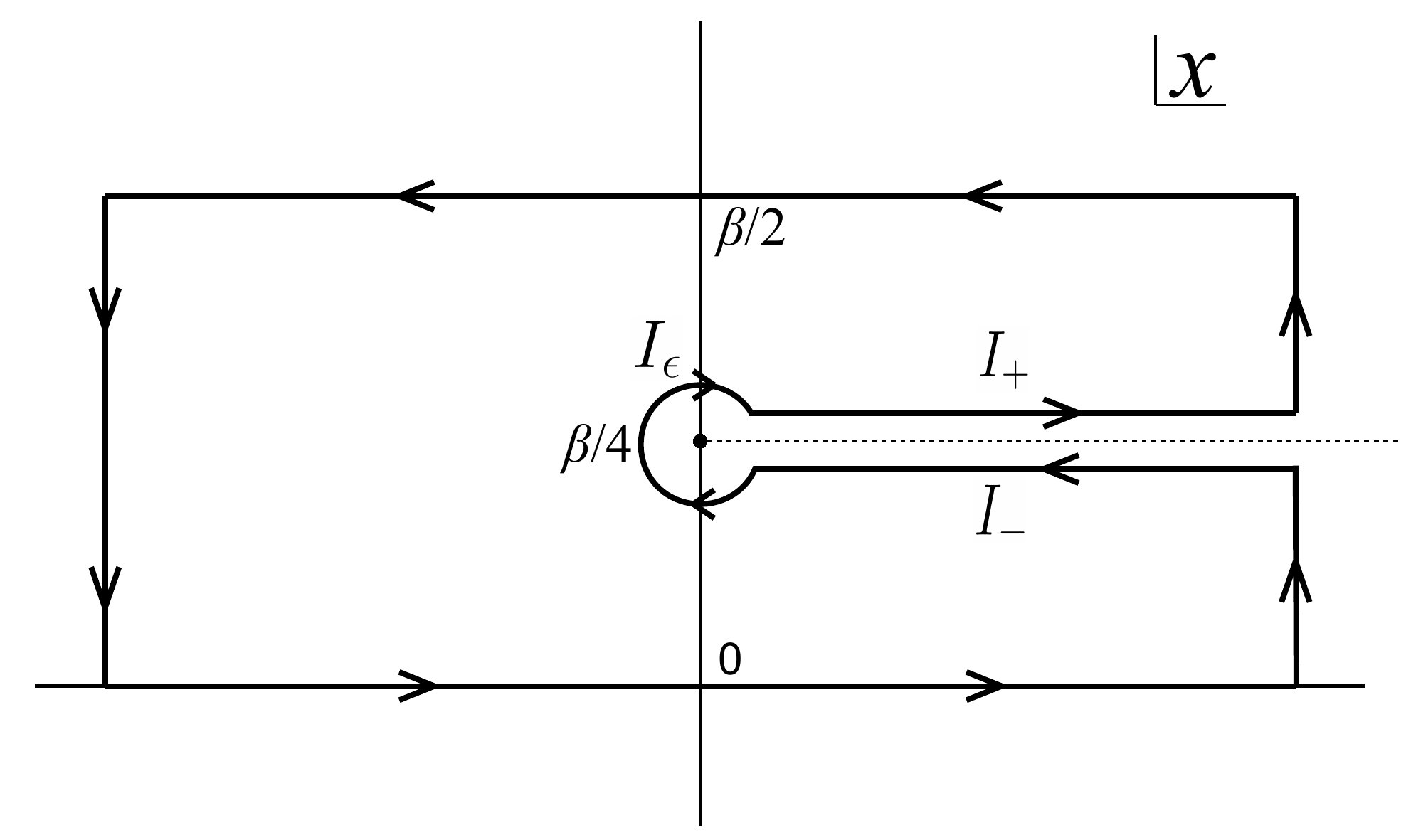}}
\caption{Contour used to calculate the reflection coefficient of a Schwarzschild BH. The singularity is a branch point and the cut extends to infinity, parallel to the real line.}
\label{fig:SchwarzContour}
\end{figure}

Next we consider cases in which there are branch points as well. It turns out \cite{basha} that generically, branch points coincide with poles of the potential, yet the leading contribution comes from the poles.
Let us illustrate this in the case of a 4D  Schwarzschild BH. In that case the tortoise coordinate is related to the radial direction in the Schwarzschild solution by
\be
x = r + 2M \log\left(\frac{r}{2M} - 1 \right).
\ee
On the complex $x$-plane, the singularity is at $x= i \beta /4$ with $\beta = 8 \pi M$. Near the singularity, the potential behaves as,
\begin{equation}\label{SchwarzPotential}
V(x) \sim -\frac14 \frac{1}{\left(x-i\beta/4\right)^2} - \frac{i}{3}\sqrt{\frac{\pi}{2 \beta}} \frac{1}{\left(x-i\beta/4\right)^{3/2}}
\end{equation}
where we kept only the two leading terms. As can be seen, there is indeed a branch point at the singularity. We chose a cut that extends to infinity and deform the contour accordingly (see figure \ref{fig:SchwarzContour}). Compared to figure \ref{fig:baseContour}, the new additions to the contour are the small circle, $I_\epsilon$, and parallel lines that run on either side of the cut, $I_\pm$. $I_\epsilon$ receives contributions only from the first term in \eqref{SchwarzPotential} and is simply $$I_\epsilon = \pi p\, e^{\frac{\beta p}{2}}.$$ That is, as expected, it behaves as a pole.

Since $V(x)$ goes to zero at infinity, in order to calculate $I_\pm$ to leading order in the large $p$ limit it suffices to treat $V(x)$ as if it behaves as in \eqref{SchwarzPotential} all the way up to infinity. Here, only the discontinuous term in \eqref{SchwarzPotential} contributes and we have,
$$
I_+ + I_- \sim p^{1/2} e^{\frac{\beta p}{2}},
$$
which is sub-leading compared to $I_\epsilon$.

In \cite{basha} we consider various black holes at the general relativity level and among other things show that the fact that contributions from branch cuts turn out to be sub-leading is generic.

\section{Classical $SL(2,\mathbb{R})_k/U(1)$ BH }

In this section we consider (\ref{uj}) for the $SL(2,\mathbb{R})_k/U(1)$ BH at the classical  (in $\alpha'$) level. Namely we are ignoring all $\alpha'$ corrections.  Thoughout the paper we set $g_s=0$, so in this section we are working at the GR (or SUGRA) level. In this case, both $V(x)$ and $R(p)$ are known exactly. Hence, we are basically verifying that the relation (\ref{uj}) works as expected.

\begin{figure}
\centerline{\includegraphics[width=0.65\textwidth]{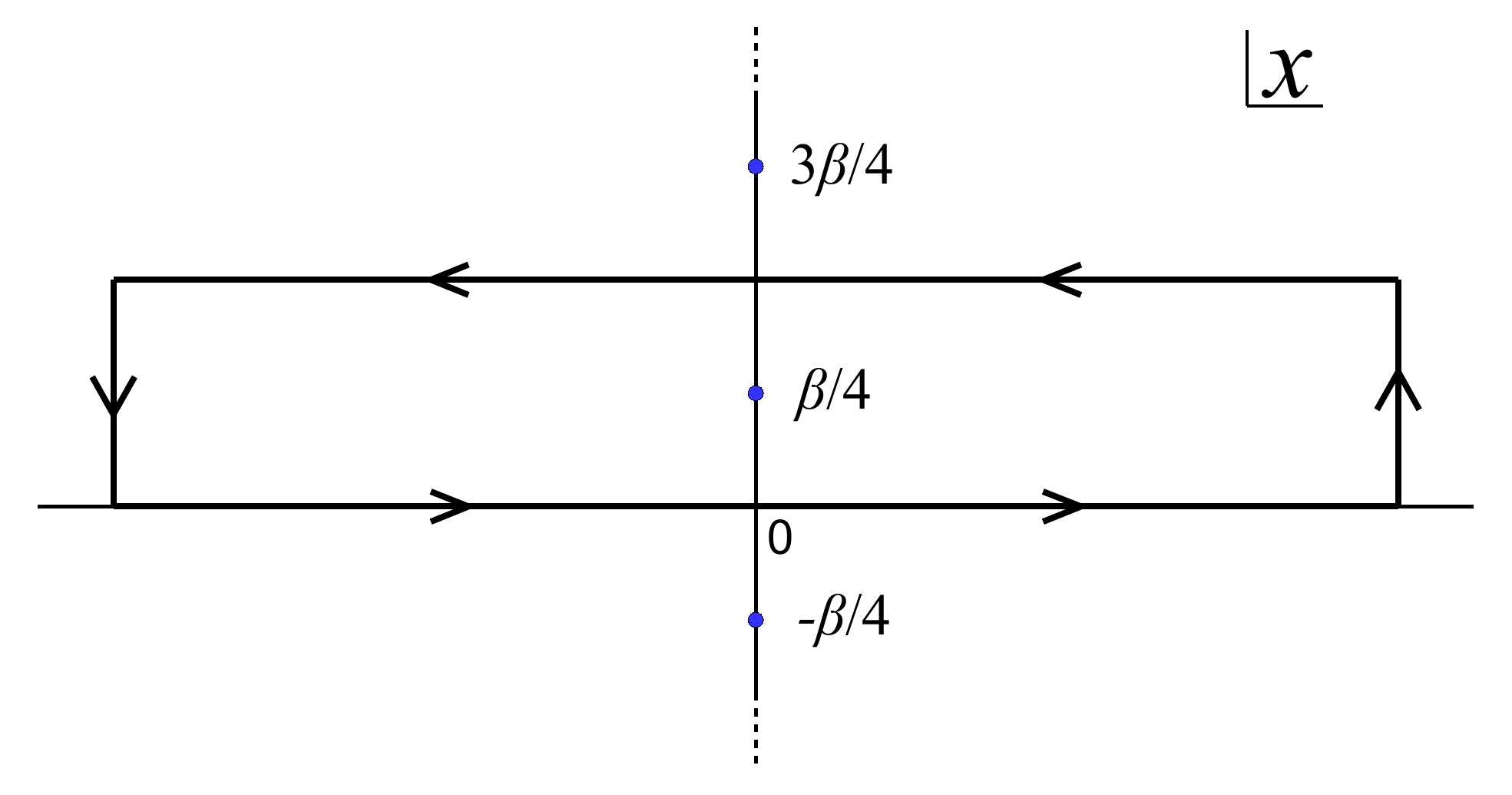}}
\caption{Poles of the potential \eqref{potentialSL2}. It has one pole at the center of the strip, $x= i \beta /4$, and infinitely many copies due the periodicity $x\to x + i \beta /2$.}
\label{fig:classSL2contour}
\end{figure}

Classically, the $SL(2,\mathbb{R})_k/U(1)$ BH is described by the background \cite{Elitzur:1991cb, Mandal:1991tz,Witten:1991yr },
\begin{equation}\label{classicalBackground}
ds^2=-\tanh^2 \left(\frac{\rho}{\sqrt{2 k}}\right) dt^2+d\rho^2  ,~~~~
e^{-2 (\Phi - \Phi_0)}= \cosh^2 \left(\frac{\rho}{\sqrt{2 k}}\right)\, ,
\end{equation}
where $\Phi$ is the dilaton and we work with $\alpha' = 2$. The wave equation in this background,
in a Schr\"{o}dinger form, has
\be \label{potentialSL2}
V(x) = \frac{1}{2k}\left(1- \frac{1}{\left(1+ e^{\sqrt{\frac2k} x}\right)^2}\right).
\ee
The potential is periodic with $\beta = 2\pi \sqrt{2k}$ and there are no branch points.
 There is, however,  a single pole in the strip at $x = \frac{i \beta}{4}$ (see figure \ref{fig:classSL2contour}). The Laurent series around the pole reads,
\begin{equation}\label{potentialSL2_NearSingularity}
V(x) =  -\frac{1}{4 \left( x - i\pi \sqrt{\frac{k}{2}} \right)^2} +\frac{1}{2 \sqrt{2k} \left(x - i\pi \sqrt{\frac{k}{2}} \right)} + \mbox{regular terms}.
\end{equation}
Thus, according to (\ref{uj}),
\begin{equation}\label{po}
R(p) \sim  e^{-\frac{\beta p }{2}},
\end{equation}
to leading order in large $p$.

The exact reflection coefficient in this case is \cite{Dijkgraaf:1991ba}
\be \label{R_classical}
R_{\text{class.}}= \frac{\Gamma(i\sqrt{2k}p)\Gamma^2\left( \frac12(1-i\sqrt{2k}p-i\sqrt{2k}\omega)\right) }
     {\Gamma(-i\sqrt{2k}p)\Gamma^2\left( \frac12(1+i\sqrt{2k}p-i\sqrt{2k}\omega )\right) }.
\ee
Using the on-shell condition and taking the large energy limit we indeed find  (\ref{po}).

\section{Perturbative (in $\alpha'$) $SL(2,\mathbb{R})_k/U(1)$ BH}

The bosonic $SL(2,\mathbb{R})_k/U(1)$ BH receives $\alpha'$ corrections already at the perturbative level. In this section we discuss different aspects of these corrections and show how they fit neatly with (\ref{uj}).

Because of the underlying $SL(2,\mathbb{R})$ structure the perturbative $\alpha'$ corrections can be computed exactly for the $SL(2,\mathbb{R})_k/U(1)$ BH background \cite{Dijkgraaf:1991ba}.
The underlying $SL(2,\mathbb{R})$ determines $L_0$ and $\bar{L}_0$ in terms of quadratic derivatives of the target space coordinates. This, in turn, determines  the effective background associated with the $SL(2,\mathbb{R})_k/U(1)$ BH.
By effective background we mean a background that involves only the dilaton and metric, such that the Klein-Gordon equation associated with it gives the $L_0$ and $\bar{L}_0$ that are determined by the $SL(2,\mathbb{R})$ structure. The effective background that takes into account perturbative  $\alpha'$ corrections, reads \cite{Dijkgraaf:1991ba}
\begin{equation}\label{per}
\begin{aligned}
ds^2 &= -\frac{k-2}{k \coth^2 \left(\frac{\rho}{\sqrt{2(k-2)}}  \right) -2} dt^2 + d\rho^2,\\
e^{-2 (\Phi - \Phi_0)} &= \frac12 \sinh \left(\sqrt{\frac{2}{(k-2)}} \rho \right) \sqrt{ \coth^2 \left(\frac{\rho}{\sqrt{2(k-2)}}  \right)- \frac2k}.
\end{aligned}
\end{equation}
We see that this background indeed has $1/k$  corrections compared to \eqref{classicalBackground}. These are naturally interpreted as perturbative $\alpha'$ corrections. In particular, as expected, for large $k$ the effective background is regular at the horizon ($\rho = 0$). Note that the inverse temperature does not receive any corrections and remains $\beta = 2 \pi \sqrt{2 k}$.

The $\alpha'$ corrections are expected to become important near the singularity. Indeed, they shift the curvature singularity from $\rho=i\pi \sqrt{k/2}$ in the classical case to
\begin{equation}
\rho_{\pm} = i \pi\sqrt{\frac{k-2}{2}} \pm \sqrt{\frac{k-2}{2}}\log \left(\frac{\sqrt{k}+\sqrt{2}}{\sqrt{k}-\sqrt{2}}\right).
\end{equation}
Note that, as expected, the shift is of order $1$ in stringy units even in the large $k$ limit. On top of the curvature singularity there is a dilaton singularity at
\begin{equation}
\rho_0 = i \pi\sqrt{\frac{k-2}{2}},
\end{equation}
that is too a stringy distance away from $ \rho_{\pm}$. So the classical singularity splits into three singularities.

\begin{figure}
\centerline{\includegraphics[width=0.75\textwidth]{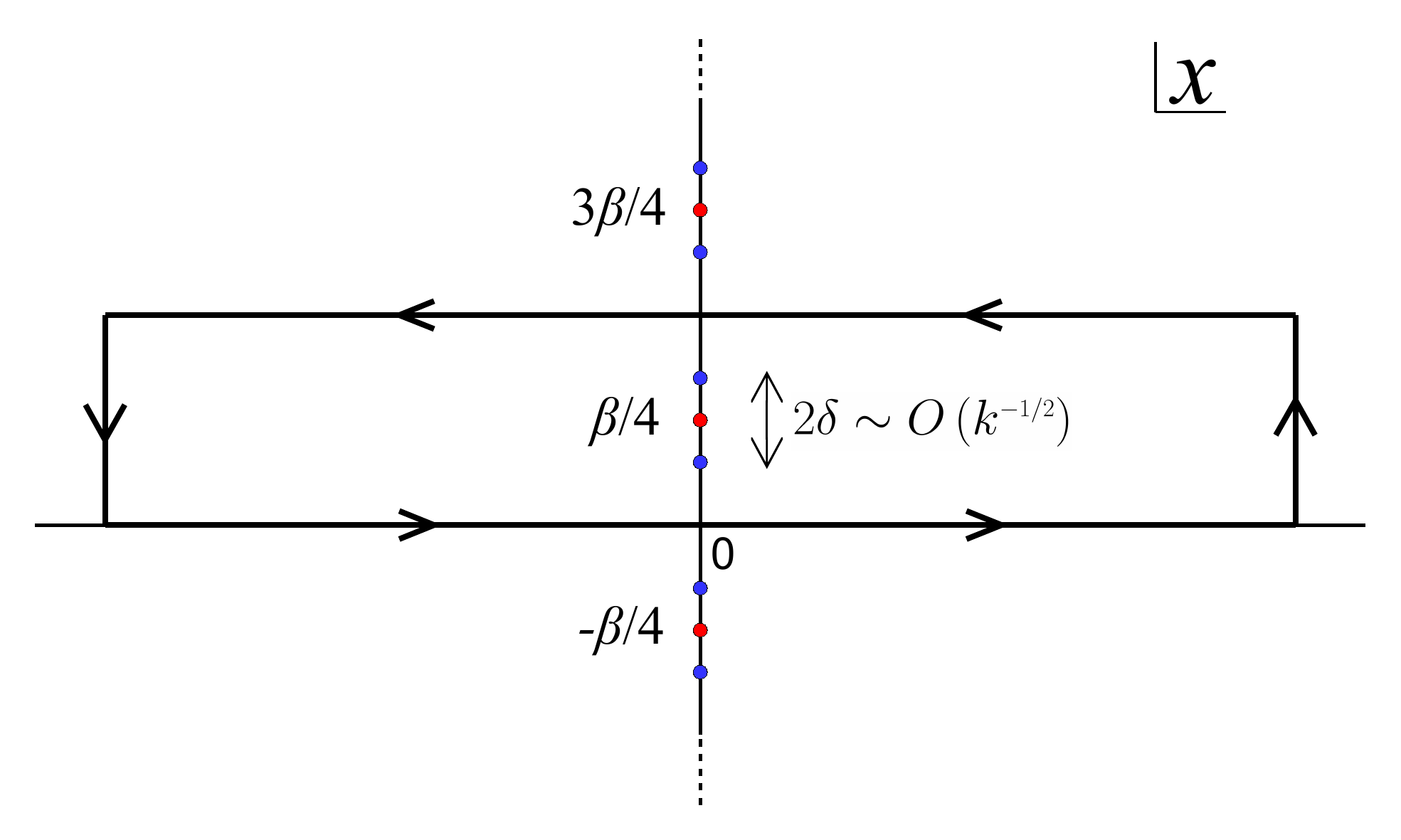}}
\caption{Poles of the potential \eqref{potentialSL2pert} in the $x$-plane. $\alpha'$ perturbative corrections splits each pole of the classical potential \eqref{potentialSL2} into three distinct ones, separated by a distance of order $1/\sqrt{k}$ (in the large $k$ limit). Poles marked in blue are curvature singularities, while those in red are dilaton singularities.
}
\label{fig:classSL2pertContour}
\end{figure}

To proceed, we have to determine the location of these singularities in the tortoise coordinate plane (see figure \ref{fig:classSL2pertContour}). The tortoise coordinate for this metric reads,
\begin{align}\label{tortoisePert}
x(\rho)&=-\sqrt{2 k} \log \left(\sqrt{\frac{k}{2}} C(\rho)+\sqrt{\frac{k}{2} C^2(\rho) -1}\right) \nonumber\\
& + \sqrt{\frac{k-2}{2}} \log \left(\frac{C(\rho) \left((k-1) C(\rho)+\sqrt{(k-2) \left(k C^2(\rho)-2\right)}\right)-1}{C^2(\rho)-1}\right),
\end{align}
where,
\begin{equation}
C(\rho)\equiv\coth \left( \frac{\rho}{\sqrt{2(k-2)}}\right).
\end{equation}
Hence, the three singularities, $\rho=\rho_0,\rho_\pm$, correspond to $C(\rho) = 0, \pm\sqrt{\frac2k}$. 

On the $x$-plane, the singularities $\rho_0, \rho_\pm$ are mapped to,
\be
x_0 = \frac{i\beta}{4},~~~~ x_\pm =  i\left(\frac{\beta}{4} \pm \delta\right), ~~~~\mbox{with} ~~~
\delta \equiv \frac{\pi}{\sqrt{2}}  \left(\sqrt{k}-\sqrt{k-2}\right).
\ee
The singularity that is closest to the line $\mIm (x) = \beta/2$, is at $x_{+}$.  The potential associated with this background is,
\begin{equation}\label{potentialSL2pert}
V(\rho)=\frac{2 (k-1) k C^6(\rho)-(k+1)^2 C^4(\rho)+6 C^2(\rho)-1}{2 C^2(\rho) \left(k C^2(\rho)-2\right)^3},
\end{equation}
where $\rho=\rho(x)$ is the inverse function of \eqref{tortoisePert}. Expanding near $x_+$ we find the leading singularity to be
\begin{equation}\label{PotentialNearPlusSingularity}
V(x) \sim -\frac{5}{36\left(x-x_+ \right)^{2}}.
\end{equation}
We note that, as for the case of the Schwarzschild BH, sub-leading terms of $V(x)$ near $x_+$ (and also $x_-$) will give rise to branch cuts. But, as discussed in section 2, in the high energy limit, it gives sub-leading contributions to $R(p)$ and can be ignored.
Therefore, in the UV we can simply use (\ref{uj}) and find the reflection coefficient to be
\begin{equation}\label{R_pert_born_largeP}
R(p)\sim  e^{-\left(\frac{\beta}{2} - 2\delta \right)p}  =  e^{-\sqrt{2(k-2)}\, \pi p}.
\end{equation}
We are  now in a position to compare this result to the exact reflection coefficient that takes the form\footnote{An additional phase $\phi(p)$, linear in $p$, was added since there is some mismatch in the definition of the reflection coefficient. The exact reflection coefficient \eqref{R_SL2_pert} (without the additional phase) is defined with respect to the coordinate $\rho$. On the other hand, the reflection coefficient calculated using the Born approximation is defined with respect to $x$. Far from the horizon, the two coordinates differ by a constant which is the cause of the phase difference.}
\be
R(p)=e^{i \phi(p)}\frac{\Gamma(i\sqrt{2(k-2)}\,p)\Gamma^2\left( \frac12(1-i\sqrt{2(k-2)}\,p-i\sqrt{2k}\,\omega)\right) }
          {\Gamma(-i\sqrt{2(k-2)}\,p)\Gamma^2\left( \frac12(1+i\sqrt{2(k-2)}\,p-i\sqrt{2k}\,\omega )\right) },\label{R_SL2_pert}
          \ee
          with
          \be2
\phi(p) = -\left(\sqrt{2(k-2)} \log (2 (k-2))+\sqrt{2k} \log \left(k-\sqrt{(k-2) k}-1\right)\right) p,
\ee
that indeed agrees with (\ref{R_pert_born_largeP}) at the UV.

\section{Exact  $SL(2,\mathbb{R})_k/U(1)$ BH}

The supersymmetric $SL(2,\mathbb{R})_k/U(1)$ BH does not receive perturbative $\alpha'$ corrections, but it does receive non-perturbative $\alpha'$ corrections.

The non-perturbative $\alpha'$ corrections
to the effective background are not known, but  the reflection coefficient, including non-perturbative $\alpha'$ corrections, is known exactly (on the sphere) \cite{Teschner:1999ug,Giveon:1999px,Giribet:2000fy, Giribet:2001ft}. Our goal in this section is to take advantage of the relation, discussed in section 2, between the reflection coefficient at the UV and the structure of the singularity to learn about some aspects of the non-perturbative $\alpha'$ corrections
to the effective background.

It is only natural to suspect that, just like the case of perturbative $\alpha'$ corrections, non-perturbative $\alpha'$ corrections will significantly affect the potential only at a stringy distance away from the semi-classical  singularity  and that in particular, the horizon will receive tiny corrections. In terms of the tortoise coordinates this means that we expect significant modifications at a distance of the order
of $1/\sqrt{k}$ away from $i\beta/4$.

At first sight, this seems to be the case since the exact reflection coefficient takes the form
\be\label{qq}
R_{\text{exact}}=\exp\left(i\theta(p)\right) R_{\text{class.}},
\ee
where $R_{\text{class.}}$ is given in \eqref{R_classical}. If the asymptote of  $\theta(p)$ was $c p$, then the location of the singularity would have been shifted by an amount $c/2$ along the (negative) real direction of $x$, as could be seen from \eqref{BornApproximation}. The reasoning above suggests that we should expect to find $c\sim 1/\sqrt{k}$.  However the calculation of Teschner \cite{Teschner:1999ug} (see also \cite{Giribet:2000fy, Giribet:2001ft,Giveon:1999px}) implies a drastically different result:
\be \label{R_stringy}
\exp\left(i\theta(p)\right)=\frac{\Gamma\left(i\sqrt{\frac2k}p\right)}{\Gamma\left(-i\sqrt{\frac2k}p\right)},
\end{equation}
which means that at the deep UV, $|p|\gg \sqrt{k}$,
\be\label{stringyPhase}
	\theta(p) \sim  \sqrt{\frac{8}{k}} \, p \log \left(\sqrt{\frac2k} \frac{|p|}{e}\right) 
	.
\ee
In \cite{Giveon:2015cma} it was shown that in the Euclidean setup this surprising result follows naturally  from the FZZ duality.

Eq. (\ref{stringyPhase}) implies that as we increase the momentum, the singularity is pushed further towards the horizon. This suggests that instead of a singular point, we should consider a cut that goes all the way to the horizon.

\begin{figure}
\centerline{\includegraphics[width=0.75\textwidth]{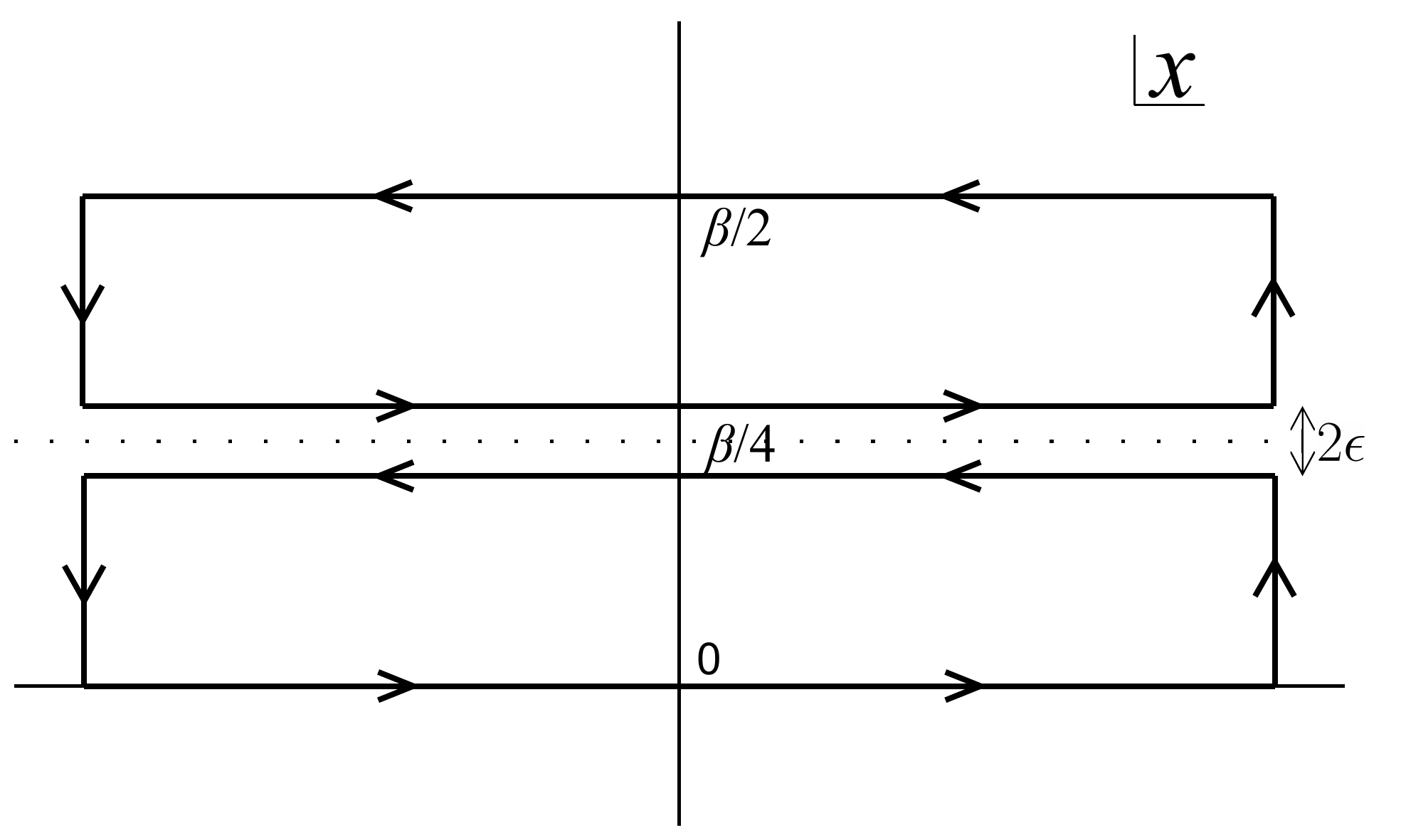}}
\caption{Contour used when including non-perturbative $\alpha'$ corrections. The rate of exponential suppression indicates that all singular features are restricted to the line $\mIm (x) = \beta/4$. The rectangle is now split in two and, since there are no other poles, the result is given by the discontinuity across the line $\mIm (x) = \beta/4$.}
\label{fig:classSL2NonPertContour}
\end{figure}

To study this in more detail we note that for $p\gg 1/\sqrt{k}$ the reflection coefficient takes the form
\be
R_{\text{exact}}(p)\sim  e^{-\frac{\beta  }{2} |p| + i \theta (p)}.
\ee
According to \eqref{uj}, it is clear that the singularity in $V(x)$ must be along the line $\mIm (x) = \beta / 4$, since this gives the correct exponential suppression, $e^{-\frac{\beta  }{2} |p|}$. To determine the nature of the singularity along this line, we split the integration contour into two, as shown in figure \ref{fig:classSL2NonPertContour}.

Assuming there is no drama at the asymptotic region and that at the horizon, the integral along each of the vertical lines vanishes, the result reads
\begin{equation}
R(p)=\frac{i}{4 p \sinh\left(\frac{\beta p}{2}\right)} \int_{-\infty}^{\infty} dx \, \mathcal{V}\left(x+\frac{i \beta}{4}\right) e^{-2 i p x},
\end{equation}
where $\mathcal{V}$ is the discontinuity of the potential in the imaginary direction
\begin{equation}
\mathcal{V} (z)\equiv  V\left(z+ i\epsilon\right)-V\left(z - i \epsilon \right).
\end{equation}
Hence, in the deep UV, we have
\begin{equation}\label{phaseDisc}
4 p\,e^{i\theta(p)} =   \int_{-\infty}^{\infty} dx \, \mathcal{V}\left(x+\frac{i \beta}{4}\right) e^{-2 i p x},
\end{equation}
which, roughly speaking, means that $\mathcal{V}$ along the line $\mIm(x)=\beta/4$ is the Fourier transform of $p\,e^{i\theta(p)}$.

Before we use  (\ref{phaseDisc}) to determine how the potential is modified due to the additional non-perturbative phase, $e^{i\theta(p)}$, we wish to illustrate how (\ref{phaseDisc}) fits neatly with the semi-classical case, discussed in section 3.
In this case we have  $e^{i \theta (p)} = 1 $ and  (\ref{phaseDisc}) gives
\begin{equation}\label{discVclassical}
 \mathcal{V}_\text{classical}\left(x+\frac{i \beta}{4}\right) = -2 i \delta ' (x).
\end{equation}
Since  the discontinuity of $1/x$ is $-2\pi i \delta (x)$, we see that the potential leading singularity is
\begin{equation}
V_\text{classical}(x) \sim
\frac{1}{\left(x - \frac{i\beta}{4}\right)^2},
\end{equation}
in accordance with \eqref{potentialSL2_NearSingularity}.

Next we consider $R_\text{exact}$. Plugging in \eqref{R_stringy}, we have\footnote{ An interesting observation is that \eqref{v_exact} happens to be the derivative of the correlator $f_{\text{non-pert}}(t)$ that appears in \cite{Ben-Israel:2015mda} (see eq. 3.5 there). At this time, we do not know whether this is pure coincidence or points to something more meaningful.},
\begin{align}\label{v_exact}
\mathcal{V}_\text{exact}\left(x+\frac{i \beta}{4}\right) = \frac{4}{\pi}\int_{-\infty}^{\infty}  \! dp \, p & \frac{\Gamma\left(i\sqrt{\frac2k }  \right)}{\Gamma\left(-i\sqrt{\frac2k } p \right)} e^{2 i p x} \nonumber \\
&= -4ik e^{-\sqrt{2k} x} J_0\left(2 e^{-\sqrt{\frac{k}{2}}x}\right),
\end{align}
where $J_0$ is the Bessel function of the first kind. This is quite different from what we had before. The discontinuities we  encountered so far were localized near the GR singularity. Here, however, $\cal{V}$ extends all the way to the horizon. In fact, near the horizon (from within the BH), $x\to -\infty$, it behaves as,
\begin{equation}\label{discontNearHorizon}
\mathcal{V}_\text{exact}\left(x+\frac{i \beta}{4}\right) \sim  -\frac{4 i k }{\sqrt{\pi }} e^{-\frac32 \sqrt{\frac{k}{2}}x} \sin \left(2 e^{-\sqrt{\frac{k}{2}}x}+\frac{\pi }{4}\right),
\end{equation}
which diverges at the horizon while wildly oscillating. Therefore, so must the potential.\footnote{ One can verify that the potential outside the horizon receives only negligible corrections.} 

We re-emphasize that in order to see this structure one has to probe the $SL(2,\mathbb{R})_k/U(1)$ BH with energies that scale like $\sqrt{k}$ \cite{Ben-Israel:2017zyi}. An external observer that probes the BH with energies only up to the string scale will conclude that its interior is the standard one up to distances of order $1$ from the singularity.

The Euclidean analog of that statement was discussed a couple of years ago \cite{Giveon:2015cma}. There it was shown that for large $k$ the way one should think about the FZZ duality is the following: at low energies (compared to $\sqrt{k}$) the proper description is in terms of the standard  cigar geometry, but high energy modes are sensitive to the sine-Liouville structure.

We expect quantum fluctuations to be quite sensitive to the difference between (\ref{discontNearHorizon}) and the standard GR potential. This implies that the stringy phase, that is at the core of the FZZ duality, should affect Hawking's original argument \cite{Hawking:1974sw} quite considerably. Similar conclusion was obtained  in  \cite{Ben-Israel:2015etg} for the  Hartle-Hawking wave function.

\section{Discussion}

In this paper we elaborated on the argument of \cite{Ben-Israel:2017zyi} that the region just behind the horizon of a {\em classical}  $SL(2,\mathbb{R})_k/U(1)$ BH is highly non-trivial.
Below we discuss various comments and questions  related to this claim.\\ \\
{\bf 1. Background vs. effective background}

Strictly speaking, we did {\em not} show that the background just behind the horizon is singular. What we showed is that the effective background, that involves only the dilaton and the metric,  is singular just behind the horizon. The effective background is the background for which the Klein-Gordon equation gives the exact reflection coefficient. It is obtained by integrating out irrelevant terms in the action. This naturally raises the possibility that it is only the effective background that is singular and not the background.

We find this possibility to be  unlikely. To have a singular effective background while having a regular background, the action should include irrelevant terms, which we denote by $X$,  with the following properties.
The background that follows from SUGRA and $X$ (and possibly other irrelevant terms)
is regular at the horizon. However, upon integrating out $X$  we get a singular background at the horizon. This implies that at the horizon $X$ couples strongly to the SUGRA action that includes the metric and the dilaton. Therefore a dilaton/graviton wave that propagates in the BH background should too couple strongly to $X$ at the horizon. In other words, such a wave will experience a non-trivial horizon.

Perturbative $\alpha'$ corrections illustrate this neatly. As reviewed  in section 4, the effective background that takes into account the perturbative $\alpha'$ corrections to the reflection coefficient is regular at the horizon. This is in accord with the fact that perturbative $\alpha'$ corrections are small at the horizon of a large BH.
This, however, also raises an interesting issue that we discuss below. \\ \\
{\bf 2. Effective description} 

Our discussion implies that there should be non-perturbative corrections (in $\alpha'$) to the SUGRA action that is sensitive to the location of the horizon. But is it possible, even in principal, to write down a term that respects diffeomorphism invariance and is sensitive to the location of the BH horizon? Naively, the answer is no. All higher order terms, such as $R_{\alpha\beta\gamma\delta} R^{\alpha\beta\gamma\delta}$, appear to be smooth, small and not sensitive to the location of the horizon.

It turns out that the situation is  more interesting \cite{Itzhaki:2004dv}. In the case of a spherically symmetric BH the following operator
\be
{\cal O}= \nabla_{\mu} R_{\alpha\beta\gamma\delta} \nabla^{\mu} R^{\alpha\beta\gamma\delta},
\ee
is sensitive to the location of the horizon as it flips sign there: it is positive outside the BH, vanishes at the horizon and negative inside the BH.  Thus, it can be considered as a horizon order parameter.

With the help of ${\cal O}$ we can write down effective actions that render the BH interior special \cite{Itzhaki:2004dv}.  For example, we can have a scalar field, $a$, whose kinetic term is ${\cal O}~
\nabla_{\mu} a \nabla^{\mu} a$. In flat space-time it vanishes and $a$ is an auxiliary field. Outside the BH, $a$ has a normal positive kinetic term, but inside the BH it has a negative kinetic term. This implies  ghost condensation \cite{ArkaniHamed:2003uy}. In such a scenario, the interior of the BH is in a different gravitational phase. The amount of ghost condensation is determined by higher order terms \cite{ArkaniHamed:2003uy} and one can construct an effective action in which the ghost condensation blows up just inside the horizon. Such an effective action might capture some of the qualitative features discussed above.

Since the $SL(2,\mathbb{R})_k/U(1)$  BH  is two dimensional it is similar in that regard to a spherically symmetric BH. Hence it is possible that operators similar to  ${\cal O}$ are generated  non-perturbatively in $\alpha'$  and  are responsible for the drama we encounter at the horizon. In fact, in this case there is a simpler operator
\be
{\cal O}_{2D}=\nabla_{\mu} \Phi \nabla^{\mu} \Phi
\ee
that flips sign at the horizon.
Whether non-perturbative $\alpha'$ corrections indeed generate terms that involve operators similar to ${\cal O}_{2D}$ or ${\cal O}$  in the effective action associated with the  $SL(2,\mathbb{R})_k/U(1)$ BH background is a fascinating question that is beyond the scope of this paper.

In non spherically symmetric situations ${\cal O}$ is not sensitive to the location of the horizon \cite{Itzhaki:2004dv}. This suggests that if there is a horizon order parameter in higher dimensions it should be non-local. This issue is related to the question discussed in the next point.
\\ \\
{\bf 3. Other Black holes}\\
In light of the results presented in this note should we expect other  BHs, such as the Schwarzschild BH,  to admit a structure at the  horizon  classically ($g_s=0)$ in string theory?
We would like to argue that the answer is no and that 
 this is a special feature of the classical $SL(2,\mathbb{R})_k/U(1)$ BH that can teach about other quantum BH. 

The reasoning is the following. In the $SL(2,\mathbb{R})_k/U(1)$ BH case the origin of the structure just behind the horizon is the extra stringy phase in the reflection coefficient.  As shown in \cite{Giveon:2015cma}, in the Euclidean case the extra stringy phase follows from the FZZ duality \cite{fzz,Kazakov:2000pm}.
Hence the wrapped tachyon field that condenses at the tip of the cigar is believed to be the seed for the extra stringy phase.
The $SL(2,\mathbb{R})_k/U(1)$ BH appears in string theory as the near horizon limit of $k$ near extremal NS5-branes \cite{Maldacena:1997cg}. Some other BHs can be obtained from it using U-duality and so we can  make some qualitative comments about these BHs too.

For example, the near horizon limit of near extremal D5-brane \cite{Itzhaki:1998dd} is obtained from the $SL(2,\mathbb{R})_k/U(1)$ BH via S-duality. The wrapped tachyon is a non-supersymmetric mode that is not protected by SUSY under S-duality. Still, it is natural to suspect that it transforms into a mode of a wrapped D1-brane located at the tip of the cigar associated with the Euclidean near extremal D5-branes. In any case, there is no reason to expect it to remain classical under S-duality. This suggests that the horizon of near extremal D5-brane is smooth at the classical level and that only  at the non-perturbative quantum level it admits a structure.

Now suppose that we compactify the D5-branes on a $T^2$ with radius $R$. Taking $R\to 0$ makes it more natural to work in the T-dual picture which takes us to a  large black hole in $AdS_5\times S^5$ \cite{Maldacena:1997re}. T-duality takes the D1-brane to a D3-brane which seems to suggest that the analog of the wrapped tachyon in this case might be a non-supersymmetric cousin of the giant gravitons discussed in  \cite{McGreevy:2000cw,Grisaru:2000zn,Hashimoto:2000zp}. At any rate, in the case of a large BH in $AdS_5$  we expect the horizon to be regular at the classical level, and to admit a structure only due to highly quantum and non-local (because of the T-duality) effects. This is consistent with the discussion in the previous item that a horizon order parameter in higher dimensions must be non-local.

To summarize, our current understanding is that the classical $SL(2,\mathbb{R})_k/U(1)$ BH appears to capture aspects of the physics  associated with quantum BHs in higher dimensions. In particular, it  admits a non-trivial structure just behind the horizon. The reason for this surprising result is not manifest in the Lorentzian setup, but in the Euclidean setup it was shown that the analog of the horizon -- the tip of the cigar -- acts too in surprising ways \cite{Kutasov:2005rr,Giveon:2013ica,Giveon:2015cma,Ben-Israel:2015mda,Giveon:2016dxe}. The origin of this is believed to be  the FZZ duality \cite{fzz,Kazakov:2000pm}. Understanding the generalization of the FZZ duality to the Lorentzian signature should considerably improve our understanding of  the results presented here.

Even without a Lorentzian generalization of the  FZZ duality progress can be made. The fact that  the classical $SL(2,\mathbb{R})_k/U(1)$ BH is an exactly solvable model makes it possible to study the non trivial horizon with the help of higher point functions. As discussed above, this could shed like on the properties of horizons associated with quantum black holes, including black holes in $AdS_5\times S^5$ backgrounds.

\section*{Acknowledgments}

We thank R. Basha, A. Giveon and  D. Kutasov for discussions.
This work  is supported in part by the I-CORE Program of the Planning and Budgeting Committee and the Israel Science Foundation (Center No. 1937/12), and by a center of excellence supported by the Israel Science Foundation (grant number 1989/14). LL would like to thank the Alexander Zaks scholarship for providing
support of his PhD work and studies.

\end{document}